\documentclass[12pt,a4paper]{article}

\usepackage{jheppub} % for details on the use of the package, please
\usepackage[T1]{fontenc} % if needed
\RequirePackage { doi } 
\usepackage{hyperref}
\usepackage{subfigure}
\DeclareMathOperator{\Tr}{Tr}

\title{\boldmath{Operator growth in SU(2) Yang-Mills theory}}
\author{Shiyong Guo }
\affiliation{Center for Theoretical Physics and Department of Physics,\\ University of California, Berkeley, CA 94720, U.S.A.}

\emailAdd{guoshy@berkeley.edu}

\abstract{Krylov complexity is a novel observable for detecting quantum chaos, and an indicator of a possible gravity dual. In this paper, we compute the Krylov complexity and the associated Lanczos coefficients in the SU(2) Yang-Mills theory, which can be reduced to a nonlinearly coupled harmonic oscillators (CHO) model. We show that there exists a chaotic transition in the growth of Krylov complexity. The Krylov complexity shows a quadratic growth in the early time stage and then grows linearly. The corresponding Lanczos coefficient satisfies the universal operator growth hypothesis, i.e., grows linearly first and then enters the saturation plateau. By the linear growth of Lanczos coefficients, we obtain an upper bound of the quantum Lyapunov exponent. Finally, we investigate the effect of different energy sectors on the K-complexity and Lanczos coefficients.}

\begin{document}
\maketitle
\flushbottom

\section{Introduction}
The study of quantum chaotic systems has gained extensive attention in the last few years. In classical mechanics, the chaos of a system is described as the sensitivity of the particle trajectory to the initial conditions. Or, in other words, the sensitivity of the position of a system in the phase space to the initial conditions. For chaotic systems, the trajectory can deviate exponentially due to small changes in the initial conditions, which is called the butterfly effect. However, quantum chaos is still difficult to be well understood. One important reason is that the concept of trajectory becomes ill-defined after considering quantum effects. One has difficulty in finding a physical observable in quantum systems to measure the extent of chaos. Despite the difficulties in measuring quantum chaos, great progress has been made in the last few years in understanding quantum chaos. To date, many methods have been developed to probe quantum chaos.

It is worth noting that the study of holography has greatly advanced the understanding of quantum chaotic systems. One of the most exciting and promising examples is the study of operator growth. In a quantum chaotic system, a simple operator becomes extraordinarily complicated after the Heisenberg time evolution. Measuring how an operator becomes more and more “complex” from a “simple” can be an important indicator in chaotic systems. Similar to the classical case, we expect the measurement to be sensitive enough to the initial conditions, and the deviation caused by different initial conditions (or rather, the complexity of the operator) grows exponentially with time.

The out-of-time-order correlator (OTOC) method has gained a lot of attention in the study of operator growth \cite{Rozenbaum:2016mmv,Hashimoto:2017oit,Nahum:2017yvy,Shen:2017kez,Fan:2016ean,Khemani:2017nda,Rozenbaum:2018sns,Pilatowsky-Cameo:2019qxt,Xu:2019lhc,Rozenbaum:2019nwn,Hashimoto:2020xfr}. One reason is that OTOC can reproduce the exponential growth in classical chaotic models. Using the OTOC method, we can generalize the Lyapunov exponent to quantum systems, i.e., the quantum Lyapunov exponent. Another important reason is that, in the context of AdS/CFT correspondence or quantum gravity, it can be an indicator of possible gravity dual\cite{Shenker_2014_1,Shenker_2014_2,Maldacena_2016}. Gedanken experiment about shock waves in the black hole background can lead to a maximum bound of Lyapunov exponent\cite{Maldacena_2016}. This Lyapunov bound is saturated in the SachdevYe-Kitaev (SYK) model \cite{Sachdev:1992fk}, which implies that the gravity duality of the SYK model may describe a quantum black hole. In addition, the operator size has been argued to correspond to the radial momentum of the particle in the gravity side\cite{https://doi.org/10.48550/arxiv.1802.01198}. The radial momentum is defined by the maximal volume of the Cauchy surface in the bulk. We will see that this has a very close relation with another important concept in quantum chaotic systems, circuit complexity.

In parallel developments, circuit complexity (or computational complexity) is another physical observable that has been given significant importance in the study of AdS/CFT and quantum chaos\cite{https://doi.org/10.48550/arxiv.1403.5695,https://doi.org/10.48550/arxiv.0804.3401,Stanford_2014,https://doi.org/10.48550/arxiv.1411.0690,Brown_2016,PhysRevLett.116.191301}. In 2014, Susskind and Stanford proposed that the circuit complexity in the boundary state is dual to the maximal volume of the Cauchy surface in bulk. Since then several proposals for the holographic complexity have been proposed, including the CA correspondence \cite{Brown_2016,PhysRevLett.116.191301} and CV2.0 correspondence \cite{Couch_2017}. To be more specific, in the study of the eternal AdS-Schwarzchild black hole, it was found that the volume of the wormhole connecting two boundaries grows linearly with time. And an important feature of computational complexity is it grows linearly in chaotic systems until the Heisenberg time $t \sim e^S$. The conjecture is the growth of the wormhole in bulk is dual to the growth of the complexity on the boundary.

Although both the studies of operator growth and circuit complexity have made exciting progress in their respective directions. However, the relation between these two similar but parallel directions has been ambiguous. An emerging concept in recent years, Krylov complexity (or K-complexity in short) \cite{Parker_2019,Barb_n_2019,Rabinovici_2021,Jian:2020qpp}, seems to bridge the gap between these two fields. It is not only a well-defined physical measure of operator growth but also has the same growth pattern as circuit complexity. In other words, there are possible geometric duals for K-complexity. We will focus on this concept in this paper. 

Krylov complexity was first put forward in \cite{Parker_2019} to measure the universal features of operator growth. Precisely, it measures the extent how an operator spread in the Krylov basis, which is constructed from the Baker-Campbell-Hausdorff formula for the Heisenberg time evolution of chosen operator. The process of constructing the Krylov basis is known as the Lanczos algorithm. When implementing the Lanczos algorithm, we obtain the so-called Lanczos coefficients $b_n$, which compactly contain the information in the Krylov basis. In the chaotic system, Lanczos coefficients are expected to show a linear growth pattern at the beginning and then go into saturation. Also, Lanczos coefficients are argued to have the fastest rate of growth in chaotic systems. In addition, the slope of the linear growth $b_n \sim a n$ of Lanczos coefficients is conjectured to give the upper bound $\lambda < 2a$ of the quantum Lyapunov exponent. This upper bound is saturated in the SYK model. The hypothesis itself is even valid beyond the semiclassical system where the Lyapunov exponent is ill-defined. From this point of view, Krylov complexity is important for finding a well-defined physical quantity to measure operator growth in quantum systems.

K-complexity has a growth pattern similar to that of circuit complexity, i.e., exponential growth at the beginning, followed by linear growth, and finally saturation. In the AdS/CFT correspondence, some possible gravitational duals of K-complexity have been investigated, e.g., the SYK model and JT gravity \cite{Jian:2020qpp}. Besides, one of the most important drawbacks of circuit complexity is that it is extremely hard to calculate circuit complexity in strongly interacting quantum field theory. While K-complexity seems to have a greater potential to be computed in quantum field theory\cite{Caputa:2021ori,Caputa:2021sib,Dymarsky:2021bjq}. It is for these reasons that calculating K-complexity in strongly interacting QFT becomes an interesting problem. In this paper, we will show the calculation of K-complexity in the SU(2) Yang-Mills theory. 

SU(2) Yang-Mills theory was shown to be chaotic in \cite{matinyan1981stochasticity}. It is expected to have a gravity dual in the large N and strong coupling limit. So, this model can be a starting point to investigate the relation between K-complexity and AdS/CFT correspondence. From another point of view, it can be recognized as a rather simple version of the Quantum chromodynamics (QCD). this model has attracted a lot of attention over the years. For these reasons, the SU(2) Yang-Mills theory and related models have attracted attention for many years.

This paper is organized as follows. In section \ref{section2}, we review how the SU(2) Yang-Mills theory can be reduced to a coupled harmonic oscillator(CHO) model by dimensional reduction, and analyze the chaotic nature of this model from both classical and quantum perspectives. In section \ref{section3}, we review the concepts of Krylov complexity and Lanczos coefficients, then summarized our approach to numerically implement the Lanczos algorithm. In section \ref{section4}, we numerically compute the Krylov complexity and Lanczos coefficients in the CHO model for different coupling constants. We found that Krylov complexity in the CHO model shows a quadratic growth in the early time stage, then enters a linear growth. By analyzing the Lanczos coefficients, we find an upper bound on the quantum Lyapunov exponent. In section \ref{section5}, we discuss the effect of different energy modes on K-complexity, and how different temperatures and energy affect Lanczos coefficients. We conclude this paper with a brief discussion in section \ref{section6}.

\section{Classical and quantum description of coupled harmonic oscillators}\label{section2}
In this section, we review how SU(2) Yang-Mills theory can be reduced to a coupled harmonic model (CHO) by dimensional reduction, following the work of \cite{Akutagawa_2020,matinyan1981stochasticity}. And then in section \ref{2.2}, we analyze the CHO model classically. In section \ref{2.3}, we review the OTOC method and quantum Lyapunov exponent in the CHO model.

\subsection{Reduction of SU(2) Yang-Mills theory}\label{2.1}
The four-dimensional SU(2) Yang-Mills theory has the following Lagrangian
\begin{equation}
    \mathcal{L} = (D_\mu \phi)^\dagger (D^\mu \phi) - V(\phi) - \frac{1}{4} F^a_{\mu \nu}F^{a \mu \nu}
\end{equation}
where the field strength $F^a_{\mu \nu} = \partial_\mu A^a_\nu -\partial_nu A^a_{\mu} + g \epsilon^{abc} A_{\mu}^b A_{\nu}^{c}$, and $D_\mu = \partial_\mu -i g A^a_\mu T^a$, $T^a = \sigma^a/2$ ($\sigma^a$ is the Pauli matrix ). Here, the index $a= 1,2,3$ and $\mu, \nu = {0,1,2,3}$. The scalar potential $V(\phi)$ is given by
\begin{equation}
    V(\phi) = \mu^2 |\phi|^2 +\lambda |\phi|^4
\end{equation}
The gauge condition we chose is $A^a_0 = 0$. Furthermore, we assume the fields satisfy the spatial homogeneousness condition.
\begin{equation}
    \partial_i A^a_j =0, \ \ \ \partial_i \phi =0
\end{equation}
Then the remaining degrees of freedom are $A^1_1, A^2_2$. Let $A^1_1=\sqrt{2} x(t), A^2_2= \sqrt{2}y(t)$, then the Hamiltonian becomes 
\begin{equation}
    H = \dot{x}^2+\dot{y}^2 + \frac{\omega^2}{4}(x^2+y^2) + g^2 x^2 y^2
\end{equation}
where we have introduced the frequency $\omega = g \langle \phi \rangle_{vac}$.

This Hamiltonian can also be recognized as a toy model of the Quantum
chromodynamics (QCD) with the colors $N_C = 2$. Then the coupling constant $g$ controls the non-linear interaction between the colors \cite{Matinyan:1981dj}, and $\omega$ denotes the mass, which is finite in the presence of the Higgs mechanism.

In classical systems, Poincaré sections and Lyapunov exponents are often used to measure the chaotic nature of a system. While in quantum mechanics, OTOC is the physical quantity that is most often used to measure the properties of chaotic systems. We will review these methods in the following sections.

\subsection{Classical analysis of coupled harmonic oscillators} \label{2.2}
N-dimensional Coupled harmonic oscillators (CHO) is a quantum mechanical model whose Hamiltonian is the following.
\begin{equation}
    H = \sum_{i=1}^{N}\left( p_i^2 + \frac{\omega^2}{4} q_i^2 \right) +\frac{1}{2}\sum_{i,j=1}^N g_{ij} q_i^2 q_j^2 
\end{equation}
where $\omega$ is the frequency of the harmonic oscillators. And $g_{ij}$ are the coupling constants, which determine the strength of the nonlinear interaction between different harmonic oscillators. 

When $g_{ij}$ is zero, the whole system is simplified to N decoupled harmonic oscillators, which is an integrable system. As $g_{ij}$ slowly increases, the non-linear term $g_{ij} q_i^2 q_j^2$ starts to dominate and the model becomes chaotic.

\begin{figure}[htbp]
\centering
\subfigure[$g_0 =0.1$]{
\includegraphics[width=5.5cm]{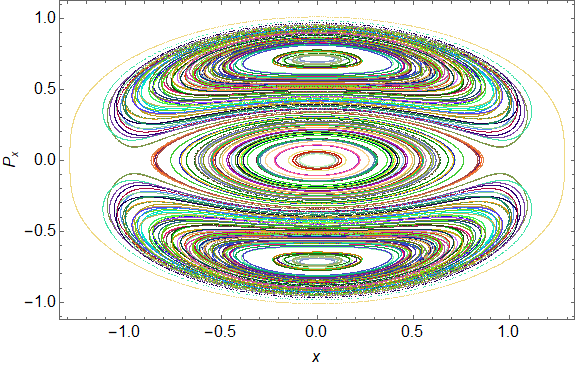}
}
\quad
\subfigure[$g_0 =0.3$]{
\includegraphics[width=5.5cm]{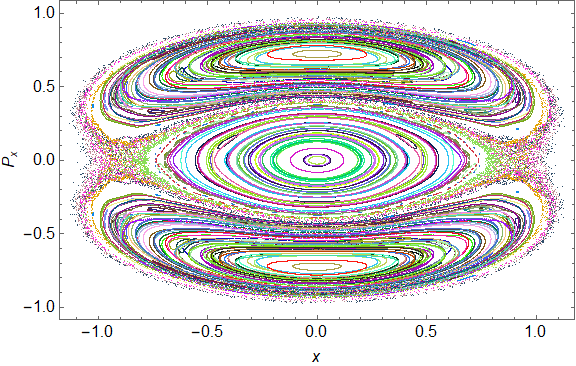}
}
\quad
\subfigure[$g_0 =0.7$]{
\includegraphics[width=5.5cm]{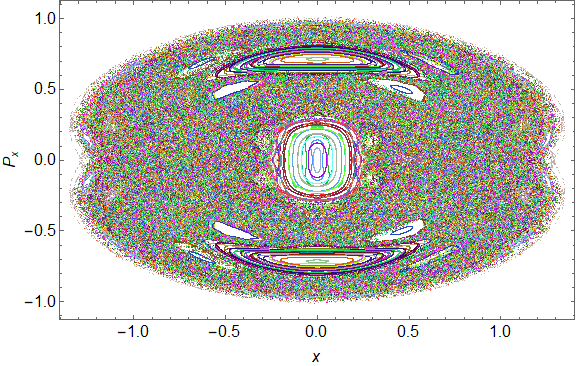}
}
\quad
\subfigure[$g_0 =1$]{
\includegraphics[width=5.5cm]{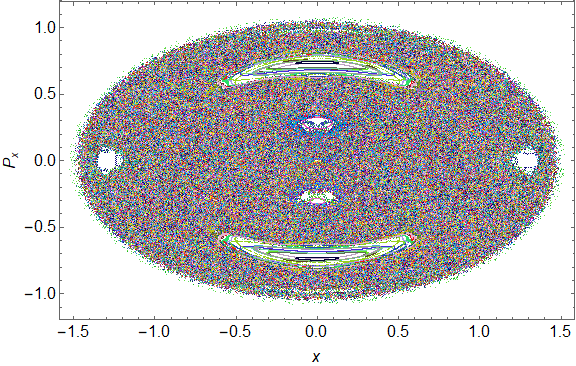}
}
\caption{Poincaré sections of the CHO model for four different coupling constants $g_0 = 0.1,0.3,0.7,1$. As the coupling constant rises, the clear orbits with strong periodicity gradually become randomly scattered points.}
\label{fig120}
\end{figure}

The Poincaré section is an intuitive way to qualitatively measure the chaotic nature of a classical system. We plotted the Poincaré sections with energy $E = 1$ for different coupling constants in Figure \ref{fig120}. It can be seen that when the coupling constant $g_0 = 0.1$, the corresponding Poincaré sections show clear orbits, which indicates the system is in a regular phase. The chaotic property of the system starts to appear when $g_0 = 0.3$. Randomly scattered points around the fixed points gradually replace the continuous trajectories. When $g_0 = 0.7$, only a small part of the Poincaré section has continuous trajectories, while most of the figure is covered by scattered points. And when the coupling constant $g_0 =1$, the Poincaré section is almost completely covered by scattered points, so the system is in the chaos phase. 

\begin{figure}[tbp]
\centering
\subfigure[E=3]{
\includegraphics[width=5.5cm]{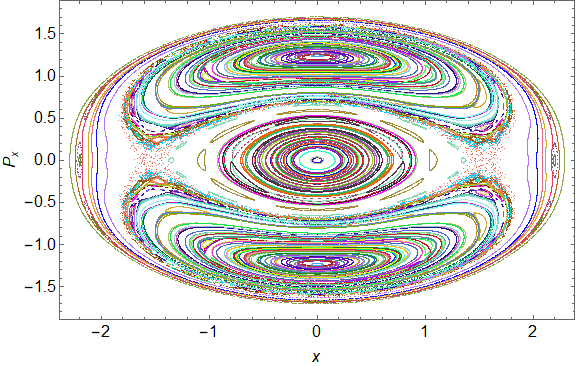}
}
\quad
\subfigure[E=5]{
\includegraphics[width=5.5cm]{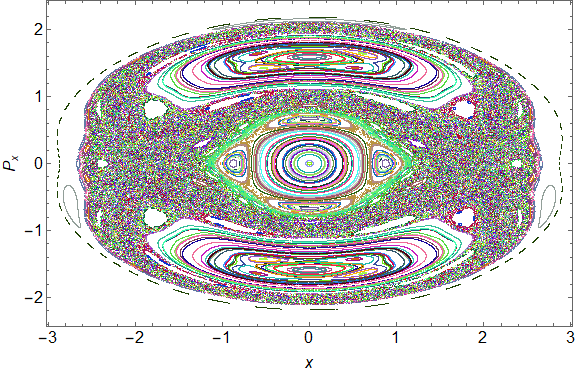}
}
\quad
\subfigure[E=7]{
\includegraphics[width=5.5cm]{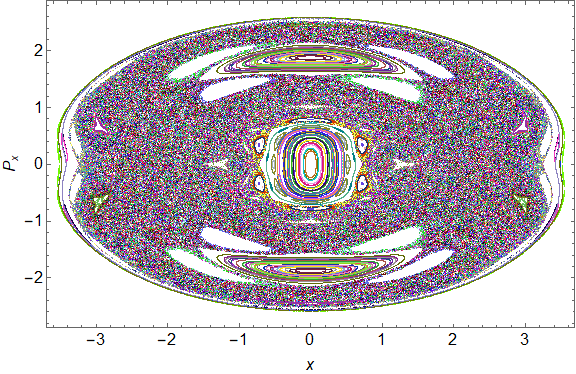}
}
\quad
\subfigure[E=15]{
\includegraphics[width=5.5cm]{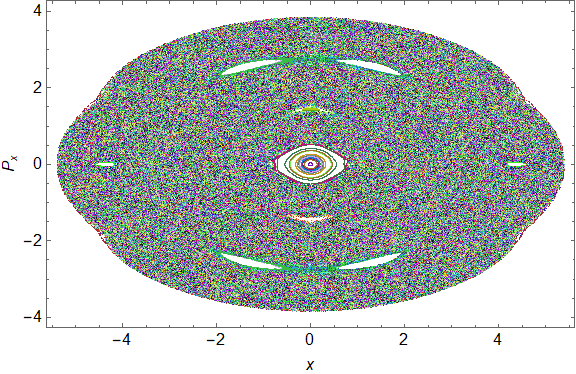}
}
\caption{Poincaré sections in the CHO Hamiltonian system for four different energy $E=3, E=5, E=7, E=15$. When the energy is low, the Poincaré section shows the characteristics of integrable systems, i.e., clear periodic orbits. As the energy increases, the system gradually enters the chaotic phase.}
\label{fig147}
\end{figure}

On the other hand, energy is also an important factor that affects the chaos in the system. We show in figure \ref{fig147} the Poincaré section at different energy $E = 1, 3, 5, 15$. We can see the transition from an integrable system to a chaotic system. At $E = 1$, the system is in regular phase. When the energy becomes $E = 3, 5$, the system transforms into a mixed phase. Finally, when $E$ rises to $15$, the system enters the chaos phase.

Another parameter that affects the chaotic property of the system is time. If the coupling constant $g_0<1$, according to the perturbative theory, when $t$ is small, the nonlinear term in the Hamiltonian is suppressed. So, the system shows the characteristics of an integrable system in the early time stage. After that, chaotic characteristics start to emerge. We analyze in section 4 the time scale of this transition.

Through the discussion in this section, we find, classically, the chaotic property of the CHO model is governed by three parameters coupling constant $g_0$, energy scale $E$, and the evolution time t. We will further discuss their effects on quantum chaotic systems in sections \ref{section4} and \ref{section5}.

\subsection{OTOC method for coupled harmonic oscillators}\label{2.3}
In classical chaotic systems, the Lyapunov exponent $\lambda_{cl}$ is defined by the exponential growth of the orbit deflection
\begin{equation}
    \Delta x(t) \simeq \Delta x(0)e^{\lambda_{cl} t}
\end{equation}
where $\Delta x(0)$ means the small differences in initial conditions, which can cause huge deviations for the orbit afterward. The corresponding exponent is defined the be the classical Lyapunov exponent $\lambda_{cl}$. This concept can be generalized to quantum systems. The quantum Lyapunov exponent is related to the out-of-time-order correlator (OTOC). In a thermal system, the OTOC is defined by 
\begin{equation}
    C_\beta(t) = -\langle [\hat{x}(t),\hat{p}]^2 \rangle_{\beta}
\end{equation}
where $\beta$ is the inverse temperature, $\langle \rangle_\beta$ denotes the thermal expectation. Here we have chosen $\hat{p}$ as the reference operator. The OTOC is expected to grow exponentially in chaotic system $C_\beta(t) \simeq C_\beta(0)e^{\lambda t}$. And the quantum Lyapunov exponent is defined to be the exponent $\lambda$.

Quantum Lyapunov exponent is an important observable to measure the chaotic structure of a system. It was conjectured that the linear growth of Lanczos coefficients gives a maximum bound on quantum Lyapunov exponent \cite{Parker_2019}. We will analyze how the Lanczos coefficients in the CHO model bound the quantum Lyapunov exponent in section \ref{section5}.

\section{Krylov complexity in coupled harmonic oscillators}\label{section3}
\subsection{Brief review of Krylov complexity}\label{3.1}

In this subsection, we briefly review the concept of operator growth and Krylov complexity. In a quantum system, the time evolution of an operator is given by
\begin{equation}
    \mathcal{O}(t) = e^{iHt}\mathcal{O}e^{-iHt}
\end{equation}
where $\mathcal{O}$ is the operator when $t=0$. We can expand it via the Baker-Campbell-Hausdorff formula
\begin{equation}
    \mathcal{O}(t) = \sum_{n=0}^{\infty} \frac{(i t)^n}{n!} \bar{\mathcal{O}}_n
\end{equation}
where $\bar{\mathcal{O}}_n$ denotes the n-th commutator of $\mathcal{O}$ with Hamiltonian
\begin{equation}
    \bar{\mathcal{O}}_0 = \mathcal{O}, \ \ \ \ \bar{\mathcal{O}}_1 = [H,\mathcal{O}], \ \ \ \ \bar{\mathcal{O}}_2 = [H,[H,\mathcal{O}]],...
\end{equation}
Alternatively, we can write in another formalism, using the well-known Liouvillian super operator. The Liouvillian super operator acts on the operator $\mathcal{O}$ by
\begin{equation}
    \mathcal{L} \mathcal{O} = [H, \mathcal{O}]
\end{equation}
In this way, we can write the time evolution of the operator as
\begin{equation}
    \mathcal{O}(t) = e^{i\mathcal{L} t}\mathcal{O}
\end{equation}

Inspired by this, we can define a set of orthonormal bases in the operator space, which is called Krylov basis. Krylov basis is constructed by imposing an iterative Gram-Schmidt orthogonalization for $\mathcal{O}_n$. This process is known as the Lanczos algorithm. However, before introducing the Lanczos algorithm, we need to determine the inner product used in the operator space. following the work of \cite{Jian:2020qpp}, we take the Wightman inner product
\begin{equation}\label{Wightman}
    (A|B) = \langle e^{H \beta/2} A^{\dagger} e^{-H \beta/2} B \rangle_\beta
\end{equation}
where the notation $\langle ... \rangle_\beta$ represents the thermal expectation of operators
\begin{equation}\label{191}
    \langle A \rangle_\beta =\frac{ \Tr (e^{-\beta H} A)}{Z},\ \ \ \  Z = \Tr(e^{-\beta H})
\end{equation}
With the definition of inner product, we can start building the orthonormal basis. The Lanczos algorithm process is as follows

\begin{enumerate}
\item Setting the starting conditions:$| \mathcal{O}_0 ) =  |\bar{\mathcal{O}}_0 )/ \Vert \bar{\mathcal{O}}_0 \Vert $, \ \ \ 
$| \mathcal{O}_1 ) =  b_1^{-1}\mathcal{L}|\mathcal{O}_0 )$,\ \ \ $b_1 = (\mathcal{O}_0\mathcal{L} |\mathcal{L}\mathcal{O}_0 )$
\item For $n \geq 2$, operator $A_n$ are constructed iteratively by:
\begin{equation}\label{151}
|A_n) = \mathcal{L} |\mathcal{O}_{n-1} ) - b_{n-1} |\mathcal{O}_{n-2} )
\end{equation}
The normalization constant $b_n$ is given by $b_n = (A_n|A_n)^{\frac{1}{2}}$. Then the basis element is constructed as follows:
\begin{equation}\label{155}
|\mathcal{O}_n) = b_n^{-1}|A_n)
\end{equation}
\item If $b_n = 0$, stop the algorithm.
\end{enumerate}

The normalization constant $b_n$ is known as the Lanczos coefficients. If we consider a system with finite-dimensional Hilbert space. Then the dimension $K$ of the space spanned by the Krylov basis is the value of n when $b_n = 0$. It was proven that Krylov space dimension satisfies $K \leq D^2-D+1$ \cite{Rabinovici_2021}, where $D$ is the dimension of Hilbert space. We will see later that this bound of the Krylov space is the reason for the saturation of Krylov complexity. However, if we consider a physical system with an infinite dimensional Hilbert space, where the Krylov space is infinite-dimensional, the condition $b_n = 0$ will never be satisfied. Accordingly, the Krylov complexity will keep growing forever. The CHO model we studied in this paper belongs to the latter one.

Once we obtained the Krylov basis, we can use it to expand the time-dependent operator $\mathcal{O}(t)$
\begin{equation}\label{159}
    |\mathcal{O}(t)) = \sum_n i^n \varphi_n(t) |\mathcal{O}_n)
\end{equation}
The Krylov amplitude $\varphi_n(t)$ is a real number and it indicates the extent to which the operator spreads on the Krylov chain. If we multiply both sides of eq(\ref{159}) by $(\mathcal{O}_n|$, we get an explicit expression for $\varphi_n(t)$.
\begin{equation}\label{amplitude}
    \varphi_n(t) = i^{-n} (\mathcal{O}_n|\mathcal{O}(t))
\end{equation}
The sum of the Krylov amplitude $\varphi_n(t)$ is conserved over time and satisfies the normalization condition. 
\begin{equation}\label{171}
    \sum_n |\varphi_n(t)|^2 = 1
\end{equation}

Since the CHO model has an infinite dimensional Hilbert space, we must make a truncation $n \leq N_{cut}$ on the Krylov basis $|\mathcal{O}_n )$ when implementing the numerical calculation. To ensure the accuracy of the numerical result, we will always keep the difference between eq(\ref{171}) and 1 no greater than $10^{-5}$. 

If we plug eq(\ref{159}) in the Heisenberg equation, $\partial_t \mathcal{O} = i[H,\mathcal{O}]$, we can obtain the following recursion relation for $\varphi_n(t)$
\begin{equation}
    \partial_t \varphi_n(t) = b_n \varphi_{n-1}(t) - b_{n+1} \varphi_{n+1}(t)
\end{equation}
From this, we can see that the time evolution of the Krylov amplitude $\varphi_n(t)$ is completely determined by the Lanczos coefficient $b_n$. This can also be seen from the matrix representation of the Liouvillian operator $(\mathcal{O}_n|\mathcal{L}|\mathcal{O}_m) = L_{nm}$
\begin{equation}
L = 
    \begin{pmatrix}
    0 & b_1 & 0 & ... & 0 \\
    b_1 & 0 & b_2 & ... & 0 \\
    0 & b_2 & 0 & ... & 0 \\
    \vdots &  & \vdots & \vdots & \vdots\\
     & ... & & ... & b_{K-1}\\
    0 & 0 & ... & b_{K-1} & 0
    \end{pmatrix}
\end{equation}

With the previous knowledge laid out, we can next introduce the definition of the Krylov complexity
\begin{equation}\label{K-complexity}
    C_K = \sum_n n |\varphi_n(t)|^2
\end{equation}
which can be recognized as the average expectation of the first-order moment of $\varphi_n(t)$ over the Krylov chain.

Krylov complexity is considered to be an important indicator of chaotic systems. For general integrable systems, it is observed that the Lanczos coefficient shows the following sub-linear growth pattern at $n\ll \ln{S}$ (S is the entropy of the system)
\begin{equation}
    b_n \sim a n^\delta
\end{equation}
where $a$ is a constant coefficient and $\delta$ satisfies $0<\delta<1$. Correspondingly, Krylov complexity satisfies polynomial growth in the early time
\begin{equation}\label{241}
    C_K(t) \sim (a t)^{\frac{1}{1-\delta}}
\end{equation}
For chaotic systems, Lanczos coefficients satisfy linear growth when n is small.
\begin{equation}
    b_n \sim a n
\end{equation}
After ending the linear growth, Lanczos coefficients enter the saturation plateau. Correspondingly, K-complexity satisfies the exponential growth in the early time stage.
\begin{equation}\label{250}
   C_K(t) \sim e^{2 a t}
\end{equation}
At this point, the coefficient $2 a$ on the exponent is considered to provide an above bound for the quantum Lyapunov exponent $\lambda$ \cite{Parker_2019}.
\begin{equation}\label{bound}
    \lambda \leq 2 a
\end{equation}

The exponential growth of the Krylov complexity continues until $t \sim \ln(S)$. From then on, the Krylov complexity becomes to grows linearly $C_K(t) \propto t$. 

The behavior of K-complexity can be traced back to the saturation of Lanczos coefficients $b_n$. For chaotic systems, the Lanczos coefficients $b_n$ reach saturation after $n \sim \ln(S)$, which leads to the linear growth of K-complexity.

For systems with finite-dimensional Hilbert space, the growth of K-complexity will continue until the Heisenberg time $t \sim e^{S}$, after which it enters the saturation plateau. The Lanczos coefficient $b_n$ starts to decrease rapidly after $n \sim e^{S}$ until it reaches zero. This operator growth behavior is conjectured to apply to all local operators in chaotic systems \cite{Parker_2019}. In the case we consider, the K-complexity will never saturate since the Hilbert space is infinite-dimensional. 

\subsection{Numerical computation of Krylov complexity}
In this subsection, we describe our approach to numerically calculate the K-complexity in CHO model. The CHO model has the following Schrödinger equation.
\begin{equation} \label{Schrödinger equation}
    -\left(\frac{\partial^2}{\partial x^2}+\frac{\partial^2}{\partial y^2}\right)\psi_n(x,y)+\left[ \frac{\omega^2}{4}(x^2+y^2) + g_0 x^2y^2 \right]\psi_n(x,y) = E_n \psi_n(x,y)
\end{equation}

By numerical calculation we can obtain a series of eigenvalues $\{E_n\}$ and eigenfunctions $\{\psi_n(x,y)\}$ of the Schrödinger equation (\ref{Schrödinger equation}). So, to implement the numerical calculation, all we have to do is to represent the Wightman inner product as a function of $ \{ E_n \}$ and $\{\psi_n(x,y)\}$.

Following the work of \cite{Akutagawa_2020}. We use $\hat{x}(t)$ as our local operator, i.e., $\mathcal{O}(t) = \hat{x}(t)$. Notice that the CHO model has the following commutation relation
\begin{equation}
  \begin{aligned}
    [H,\hat{x}] = -2ip_x, \ \ \ \ [H,\hat{y}] = -2ip_y
  \end{aligned}    
\end{equation}

Therefore, all local operators in the CHO model can be represented as functions of $\hat{x}$ and $H$. It is worth noting that the commutation relation is independent of how many oscillators are in the model. So, this computational method can be used in an arbitrary dimensional CHO model. However, in this paper, we will only focus on the two-dimensional CHO model.

In the Lanczos algorithm, we will only do the inner product among the Krylov basis $\{\mathcal{O}_n\}$. So we just need to express the inner product of the Krylov basis $\{\mathcal{O}_n\}$ as a function of $ \{ E_n \}$ and $\{\psi_n(x,y)\}$. According to (\ref{151}) and (\ref{155}), $\mathcal{O}_n$ can be expressed as a polynomial of $\hat{x}$ and $H$, where the order of $\hat{x}$ in each term is 1 and the order of $H$ is at most $n$.

From (\ref{Wightman}), the Wightman inner product can be expressed as
\begin{equation}
    (A|B) = \frac{1}{Z} \Tr \left( e^{-H \beta/2} A^\dagger e^{-H \beta/2} B \right)
\end{equation}
After inserting the completeness condition $\sum_l |l\rangle \langle l|=1$, we get
\begin{equation}
    (A|B) = \frac{1}{Z} \sum_{m,l} e^{-\frac{\beta}{2}(E_m+E_l)}A^\dagger_{ml} B_{lm}
\end{equation}
where we introduced $A_{ml} = \langle m | A | l \rangle$. Since we only need to consider the Wightman inner product among $\{|\mathcal{O}_n)\}$, and $|\mathcal{O}_n)$ can be expressed as a polynomial of $\hat{x}$ and $H$. Therefore, we have
\begin{equation}\label{234}
\begin{aligned}
    (\mathcal{O}_n)_{lm} &= \sum_{k=1}^n \sum_{a+b=k} C_{ab} E_l^a E_m^b \langle l | \hat{x} | m \rangle\\
    (\mathcal{O}_n)^{\dagger}_{ml} &= \sum_{k=1}^n \sum_{a+b=k} C^{\prime}_{ab} E_m^a E_l^a \langle m | \hat{x} | l \rangle
\end{aligned}
\end{equation}
where $C_{ab}$ and $C^{\prime}_{ab}$ are constant coefficients. According to eq(\ref{234}), we only need to numerically integrate $\langle m| x | l \rangle$ to get the value of the Wightman inner product. Through this method, we can numerically compute the K-complexity and Lanczos coefficients.

Our numerical approach can be summarized as follows

\begin{enumerate}
    \item Numerically solve the Schrödinger equation and obtain the corresponding energy eigenvalues $\{E_n\}$ and eigenfunctions $\{\psi_n(x,y)\}$.
    \item For different $m$ and $l$, compute $\langle m | \hat{x} | l \rangle = \int dx dy \ \psi_n^{\dagger}(x,y) x \psi_n(x,y)$ numerically.
    \item Implement the Lanczos algorithm by writing the Wightman inner product involved in each step as a function of $\langle m | \hat{x} | l \rangle$, then plug in $\langle n | \hat{x} | m \rangle$ obtained by step 2.
    \item Through the Lanczos algorithm, obtain Krylov basis $|O_n)$ and Lancaos coefficient $|b_n)$.
    \item Calculate the Krylov amplitude $\varphi_n$ by (\ref{amplitude}). The corresponding K-complexity is obtained by plugging $\varphi_n$ into (\ref{K-complexity}).
\end{enumerate}

\subsection{Energy spectrum and spectral form factor}
\begin{figure}[htbp]
\centering
\includegraphics[width=7cm]{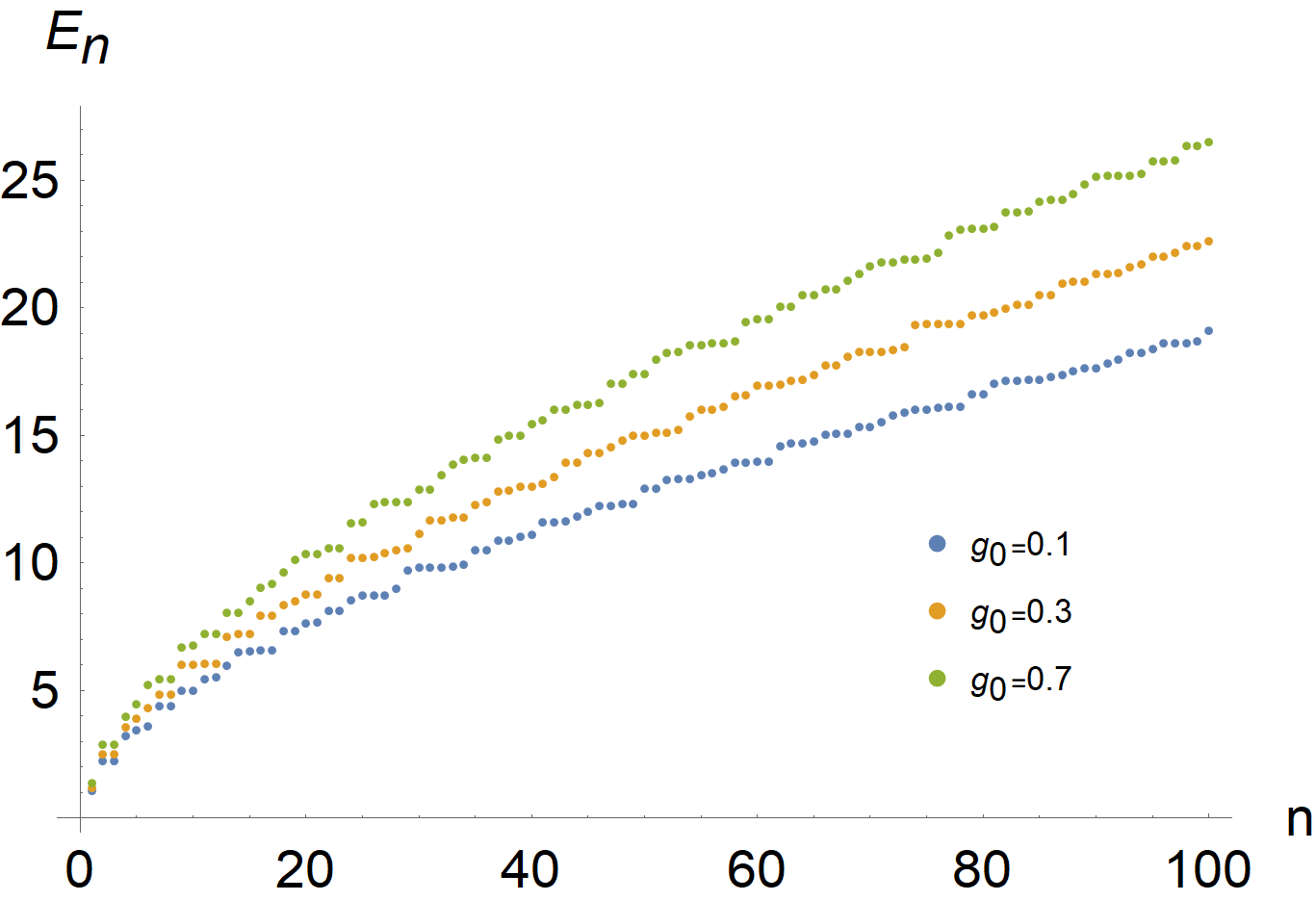}
\caption{}
\label{energyspectrum}
\end{figure}

Before moving to the calculation of the K-complexity, we show in this section the energy spectrum as well as the spectral form factor of the CHO model. Figure \ref{energyspectrum} shows the energy spectrum of the CHO model for different coupling constants from the numerical calculation of the Schrödinger equation (\ref{Schrödinger equation}).

\begin{figure}[htbp]
\centering
\subfigure[]{
\includegraphics[width=5.5cm]{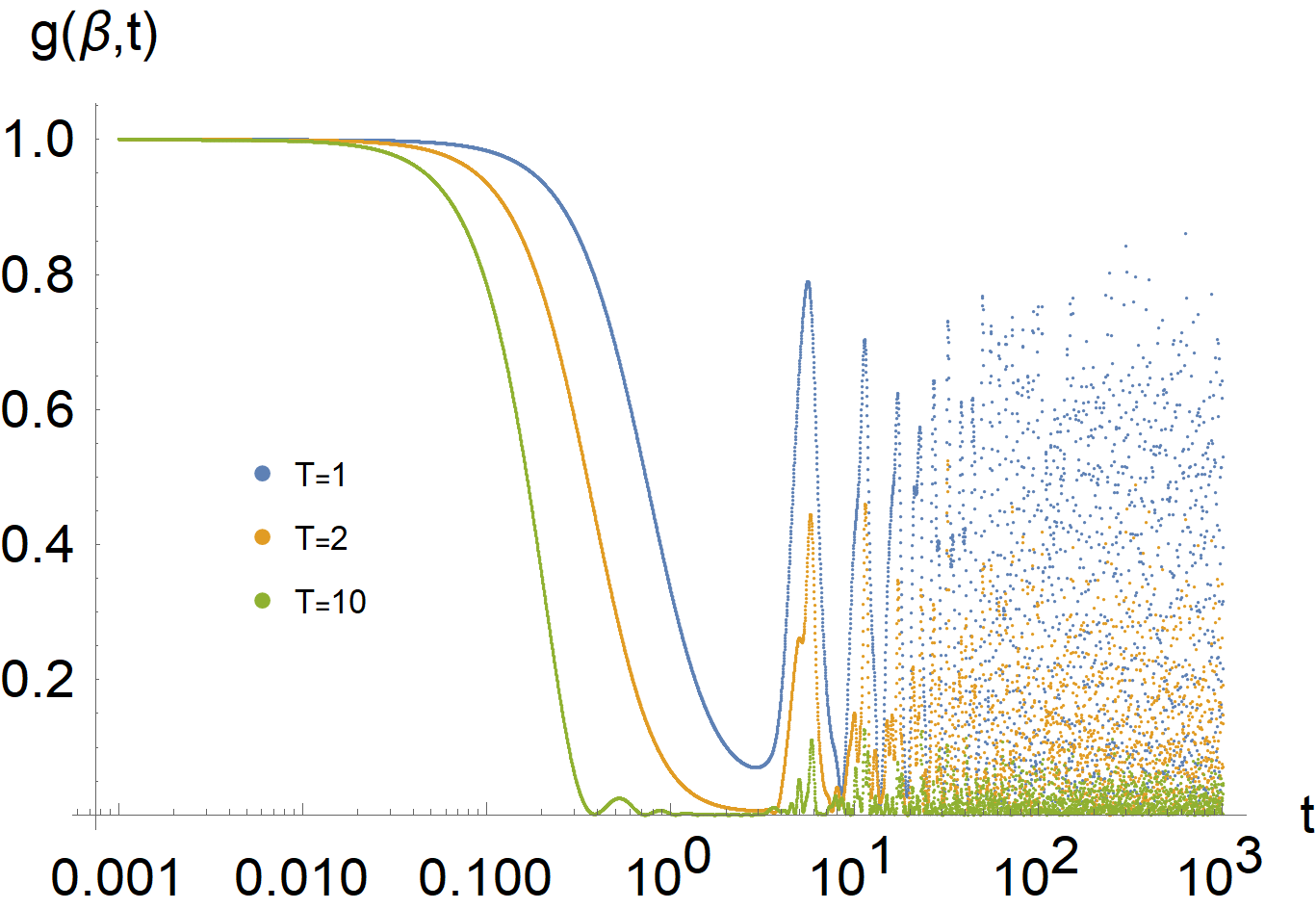}
%\caption{fig1}
}
\quad
\subfigure[]{
\includegraphics[width=5.5cm]{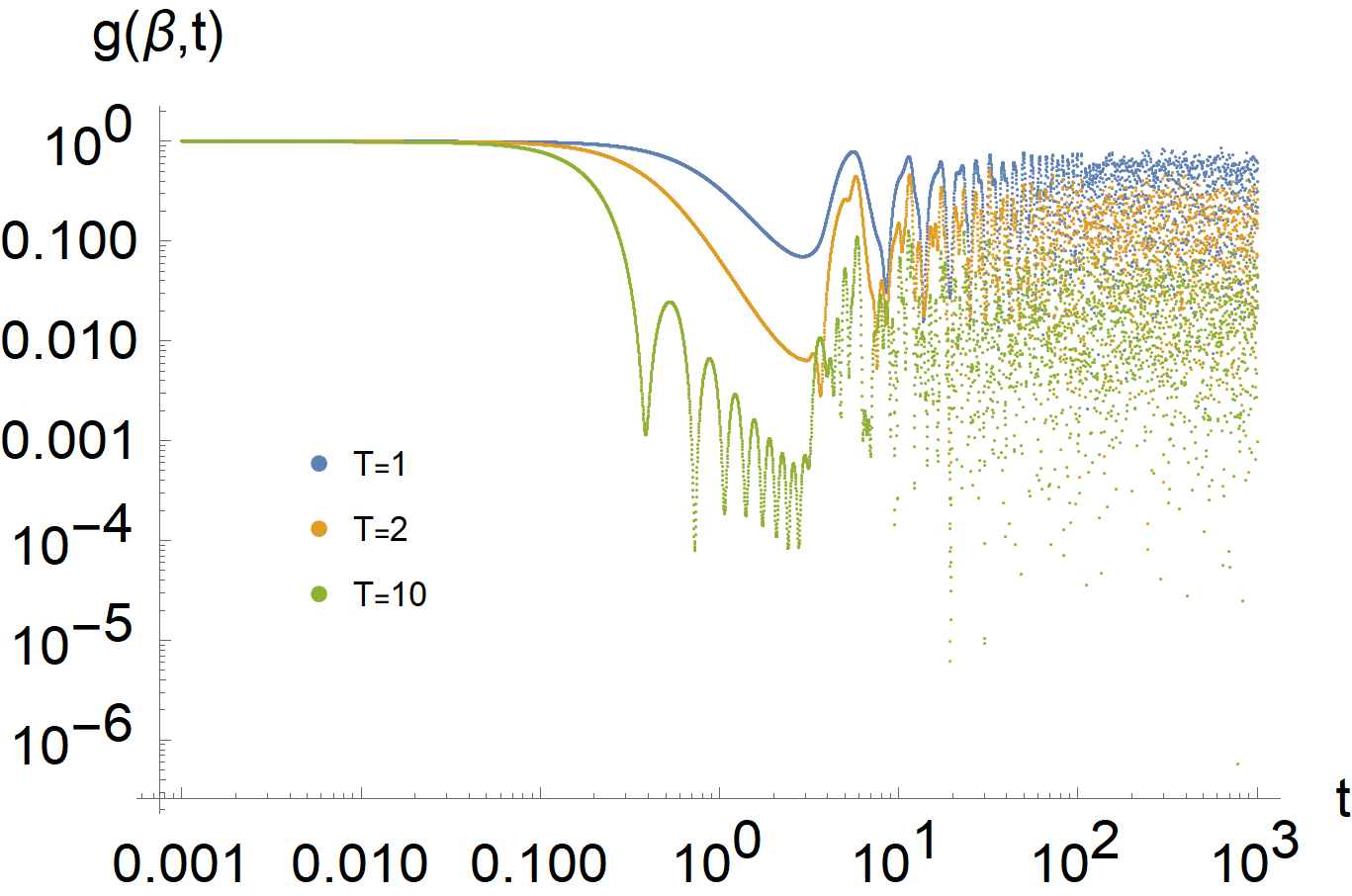}
}
\caption{Spectral form factor $g(\beta, t)$ as a function of time, for $g_0 = 0.1$ and three different temperatures $T=1,T=2,t=10$. In the early stage, the spectral form factor holds 1. After that, it drops to a dip and then bounces back to a plateau with erratic fluctuations. The fluctuation in the spectral form factor reflects the discrete eigenvalues of the system.}
\label{spectralformfactor}
\end{figure}

Another important observable in chaotic systems is the spectral form factor $g(\beta,t)$. Its behavior can show many characteristics of chaotic systems. And in recent developments it has been related to the information loss problem and AdS/CFT correspondence \cite{Dyer:2016pou,Cotler:2016fpe,Cotler:2017jue}. The spectral form factor $g(\beta,t)$ is defined by
\begin{equation}
    g(\beta,t) \equiv \frac{|Z(\beta + it)|^2}{Z^2(\beta)}
\end{equation}
where we have analytically continued partition function $Z(\beta)$ in eq(\ref{191}) into the complexity plane. In figure \ref{spectralformfactor} we show the spectral form factor evolution with $t$ for coupling constant $g=0.1$ under different temperatures $T = 1, T=2, T=10$.

At low temperatures, $g(\beta,t)$ seems to first decrease to the dip, and then bounce back to a certain value with erratic fluctuations. At high temperatures, $g(\beta,t)$ first drops and then rises to a plateau with a relatively small value. We find the spectral form factor is very much reminiscent of that from the random matrix theory \cite{Dyer:2016pou}, which implies that there may be a random matrix description for the SU(2) Yang-Mills theory.

\section{Krylov Complexity in chaotic transition}\label{section4}
In this section, we numerically compute the behavior of Krylov Complexity and Lanczos coefficient in the CHO model using the method we presented in the previous section. In section \ref{4.1}, we analyze the time scale for the system's chaotic property to start to emerge as well as the time scale for the K-complexity to grow linearly. In section \ref{4.2}, we show the growth of Lanczos coefficients with different coupling constants. We found that the Lanczos coefficient grows linearly at the beginning and then goes into saturation. In section \ref{4.3}, we present the result for the time evolution of Krylov complexity. In this section, we will fix the inverse temperature $\beta =1$.

\subsection{Time scale for chaotic transition}\label{4.1}
In the CHO model, K-complexity has two different growth patterns in the early and late time evolution. This is because, according to the BCH formula, at a very short time t, we can treat the nonlinear interaction $g_0 x^2 y^2$ in the Hamiltonian as a perturbative term. And the corresponding operator evolution can be approximated as
\begin{equation}
e^{i H t} \hat{x} e^{i H t} \sim (1+ i g_0 \hat{x}^2 \hat{y}^2 t) e^{i H_0 t} \hat{x} e^{i H_0 t} (1- i g_0 \hat{x}^2 \hat{y}^2 t) 
\end{equation}
where $H_0$ is the non-perturbative Hamiltonian. We want to find a time scale when the contribution from the non-linear term becomes important, i.e., the chaotic property of the system begins to emerge.

We assume that if the correction of the non-linear term for energy exceeds $10\%$, the effect of the coupling term becomes non-negligible. In other words, when the expectation value of $g_0 \hat{x}^2 \hat{y}^2 t$ under the ground state $| \Omega \rangle = \mathcal{N} e^{- \beta E_n} |n \rangle$ reaches $10\% E$ (E is the energy of the system and $\mathcal{N}$ is the normalization constant), the system starts to become chaotic.

By calculation, we can find that the expectation of the non-linear term $g_0 \hat{x}^2 \hat{y}^2 t$ is roughly

\begin{equation}
    \langle \Omega | g_0 \hat{x}^2 \hat{y}^2 t |\Omega \rangle \sim g_0 t
\end{equation}

If we choose $g_0 = 0.1$, then our chaotic transition time is roughly $t_0=1$. As we will show in the next sections, our estimate here matches our numerical results.

Another different time scale is the scrambling time $t_*$ at which the K-complexity starts to grow linearly. As stated in early sections, $t_* \sim \ln(S)$. The Shannon entropy $S_H$ of our system can be computed by
\begin{equation}
    S_H = -\sum_n \mathcal{N}^2 e^{-2 E_n} \ln(\mathcal{N}^2 e^{-2 E_n})
\end{equation}
Through numerical calculation, we find the scrambling time $t_*$ is roughly around $t\sim1$. We find that this is very close to the chaotic transition time $t_0$ we calculated. We will see in the following sections because the non-linear term of the system is suppressed in the early time, K-complexity in early growth satisfies not exponential growth but power-law growth. And, around the chaotic transition time $t_0$, K-complexity shifts directly from power-law growth to linear growth.

\subsection{Lanczos coefficients and upper bound for quantum Lyapunov exponent}\label{4.2}
Lanczos coefficients are an important indicator in chaotic systems. First, the growth of Krylov Complexity is captured by the Lanczos coefficients. The linear growth of Lanczos coefficients corresponds to the exponential growth of K-complexity in the early time. Generally, in chaotic systems, the Lanczos coefficients grow linearly at the beginning, which is called Lanczos ascent. After the linear growth, it enters the saturation plateau. The K-complexity keeps growing linearly while the Lanczos coefficients are on the plateau. Besides, the linear growth of the Lanczos coefficients $b_n \sim a n $ is expected to give a maximum bound $ \lambda < 2 a$ for the quantum Lyapunov exponent.

\begin{figure}[htbp] 
\centering
\includegraphics[width=0.7\textwidth]{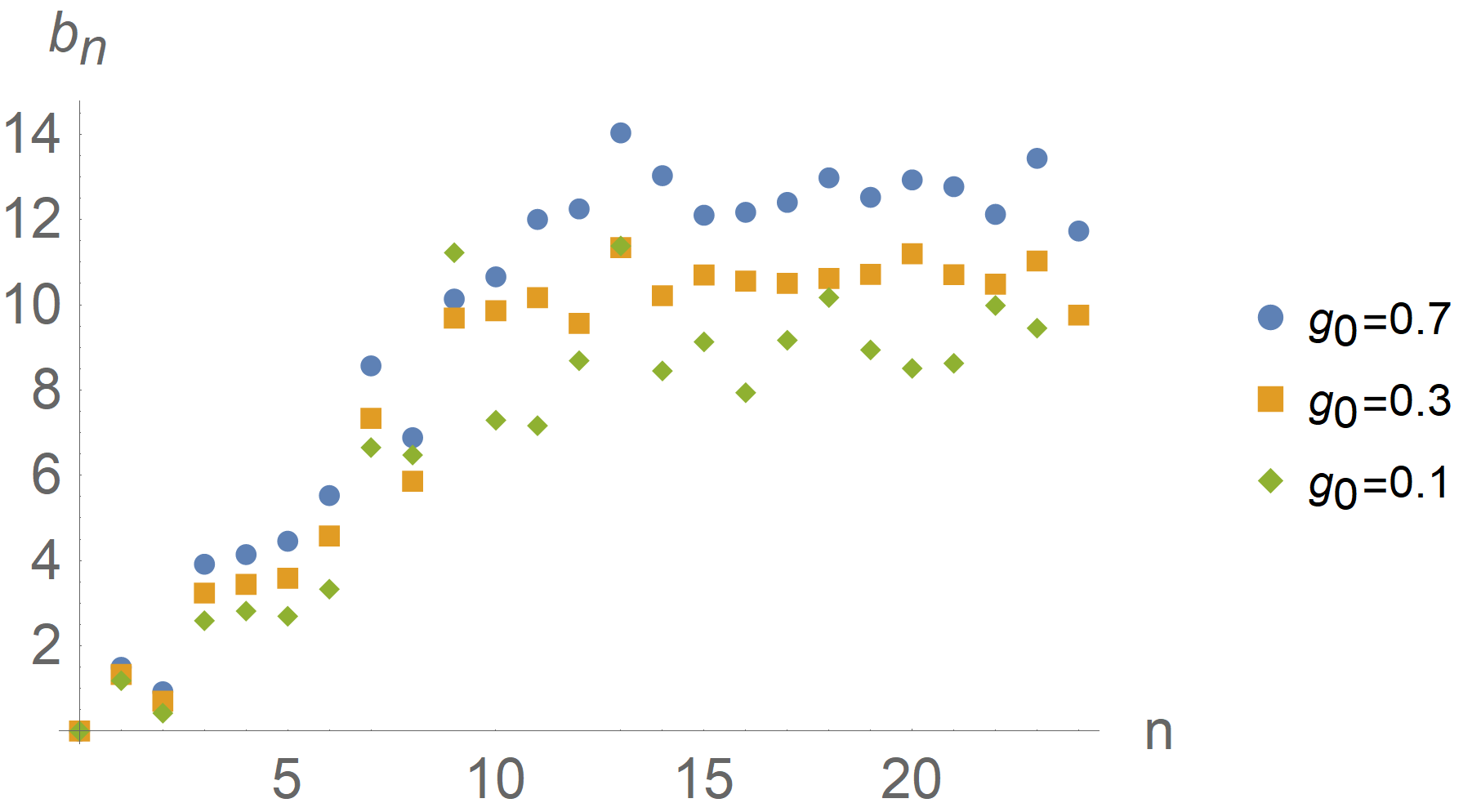} 
\caption{Lanczos coefficients for different coupling constants $g_0 = 0.1, 0.3, 0.7$ at temperature $\beta =1$. In the beginning, the Lanczos coefficients $b_n$ increase linearly with $n$. And after about n = 12, Lanczos coefficients enter the saturation plateau and fluctuate around a constant value.} 
\label{fig335} 
\end{figure}

We show in figure \ref{fig335} the Lanczos coefficients $b_n$ for different coupling constants. From the figure, we can see that the Lanczos coefficients rise approximately linearly when n is small. Around $n \sim 12$, the Lanczos coefficients go into saturation, which ensures an asymptotic linear growth of K-complexity. The system we consider has a Hilbert space with infinite dimension, so the Lanczos coefficient stays on the plateau and will not drop to zero. The effect of different coupling constants on the Lanczos coefficient is mainly to change the saturation value of Lanczos coefficients. The greater the coupling constant, the larger the saturation value of Lanczos coefficients. It can be translated to the behavior of K-complexity, i.e., the larger coupling constant will lead to a steeper linear growth for K-complexity.

Notice that there is an oscillation phenomenon in the Lanczos coefficients. One possible explanation for this oscillation is that the Schrödinger equation of Kyrlov amplitude allows the following solution\cite{Yates:2020lin,Yates:2020hhj,Bhattacharjee:2022vlt}
\begin{equation}
    b_n = f(n)+ (-1)^{n} \Tilde{f}(n)
\end{equation}
The asymptotic behavior of the oscillatory function $\Tilde{f}(n)$ can be derived from the auto-correlation function $C(t) = \varphi_0(t)$. In general, if the auto-correlation function is power-law decaying $C(t) \sim t^{-b}$, then the oscillation of the corresponding Lanczos coefficient satisfies a logarithmically decay $\Tilde{f}(n) \sim \ln (n)^{-b}$\cite{Bhattacharjee:2022vlt}.

According to (\ref{bound}), by computing the slope of the linear growth of Lanczos coefficients $b_n$, we can obtain the above bound of the quantum Lyapunov exponent. 

\begin{equation}
    \lambda_{0.1} \leq 1.57, \ \ \
    \lambda_{0.3} \leq 1.79, \ \ \ 
    \lambda_{0.7} \leq 2.09
\end{equation}
The subscript of $\lambda$ is the corresponding coupling constant.

\subsection{K-complexity growth in the chaotic transition}\label{4.3}
Through the perturbative analysis in section \ref{4.1}, we know that the non-linear term of the Hamiltonian can be neglected when $t$ is small enough. Therefore, the system can be considered as two decoupled harmonic oscillators, which is a system with the Heisenberg-Weyl symmetry. According to \cite {Caputa:2021sib}, a system with the Heisenberg-Weyl symmetry has a quadratic growing K-complexity. 

\begin{equation}
    K \sim t^2
\end{equation}

\begin{figure}[htbp]
\centering
\subfigure[]{
\includegraphics[width=5.5cm]{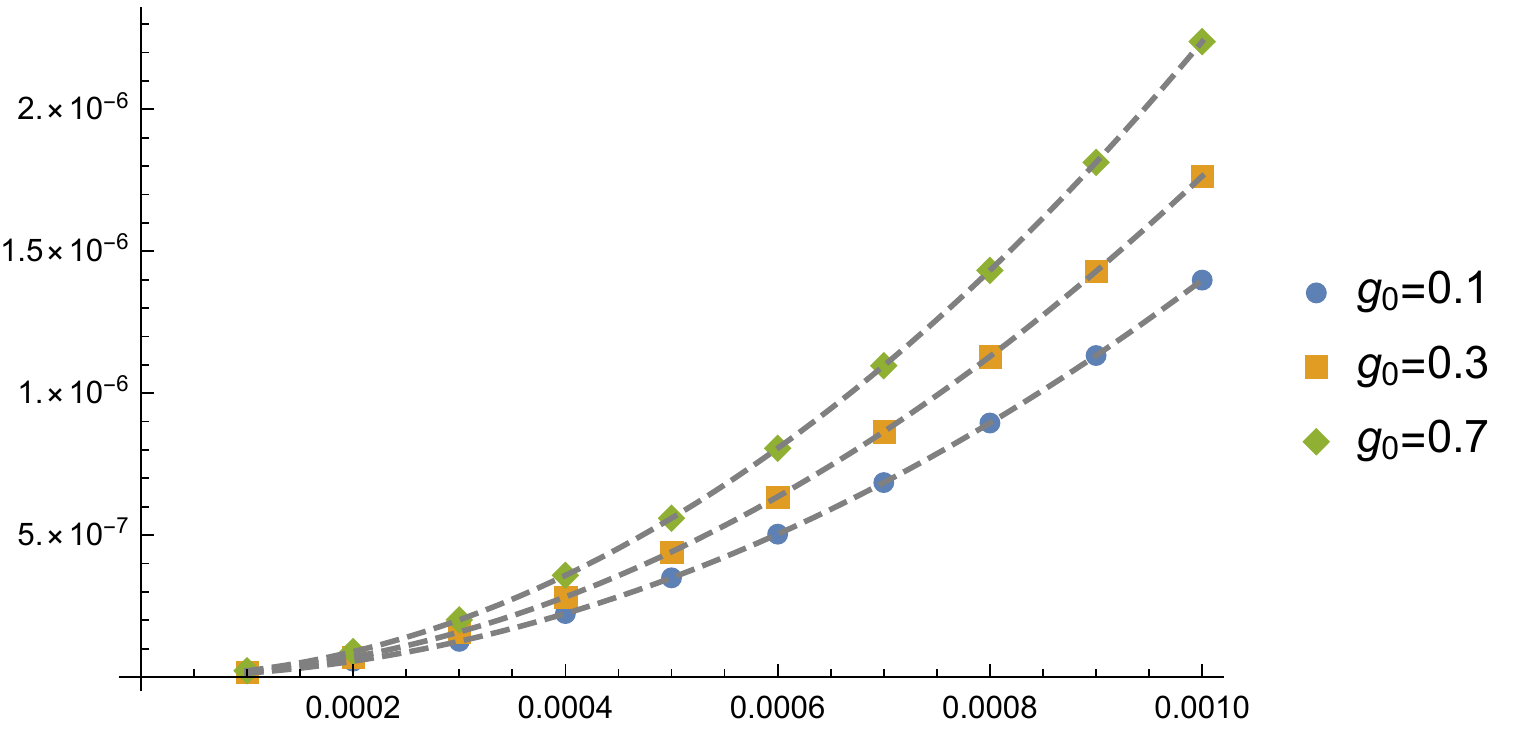}
}
\quad
\subfigure[]{
\includegraphics[width=5.5cm]{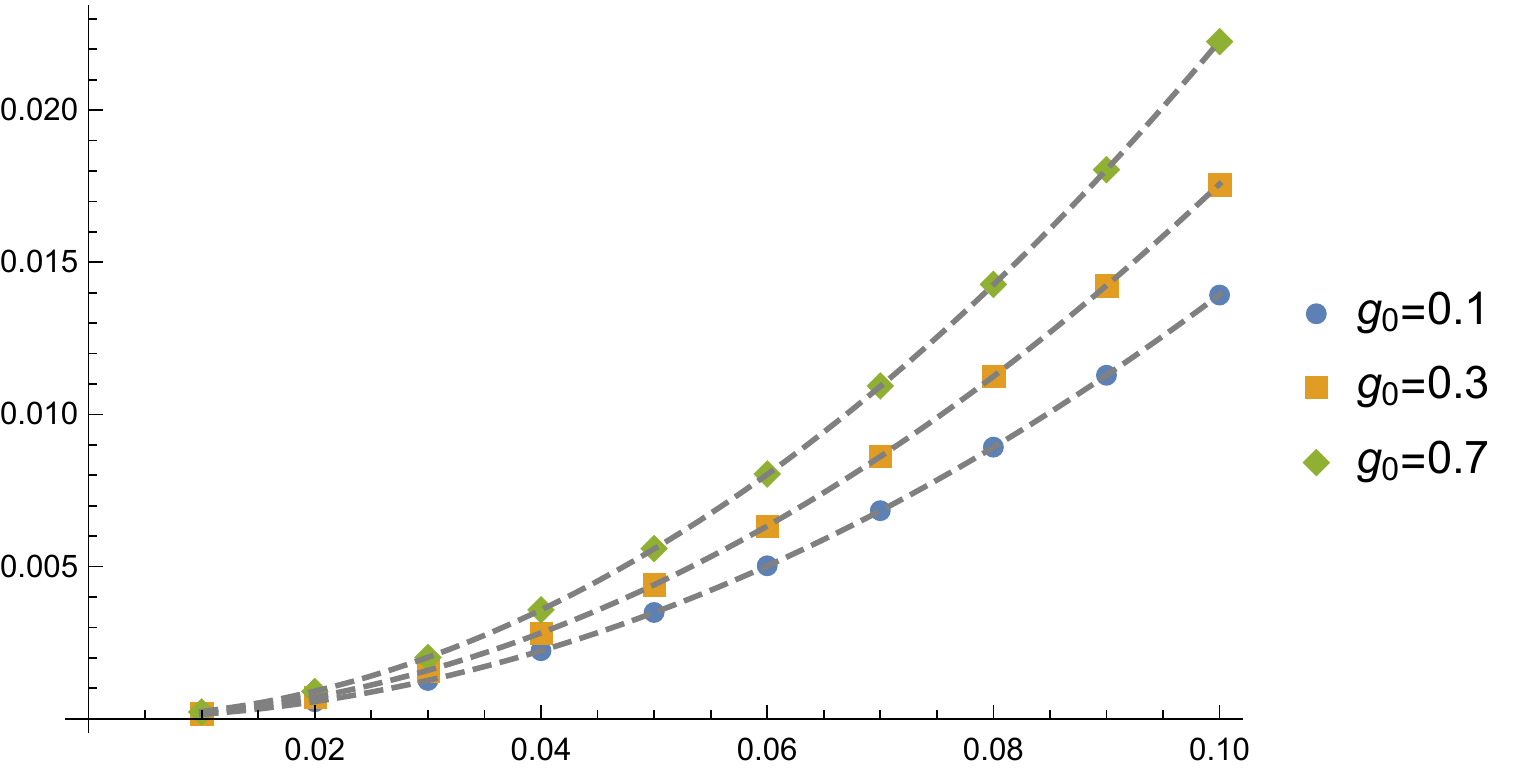}
}
\quad
\subfigure[]{
\includegraphics[width=5.5cm]{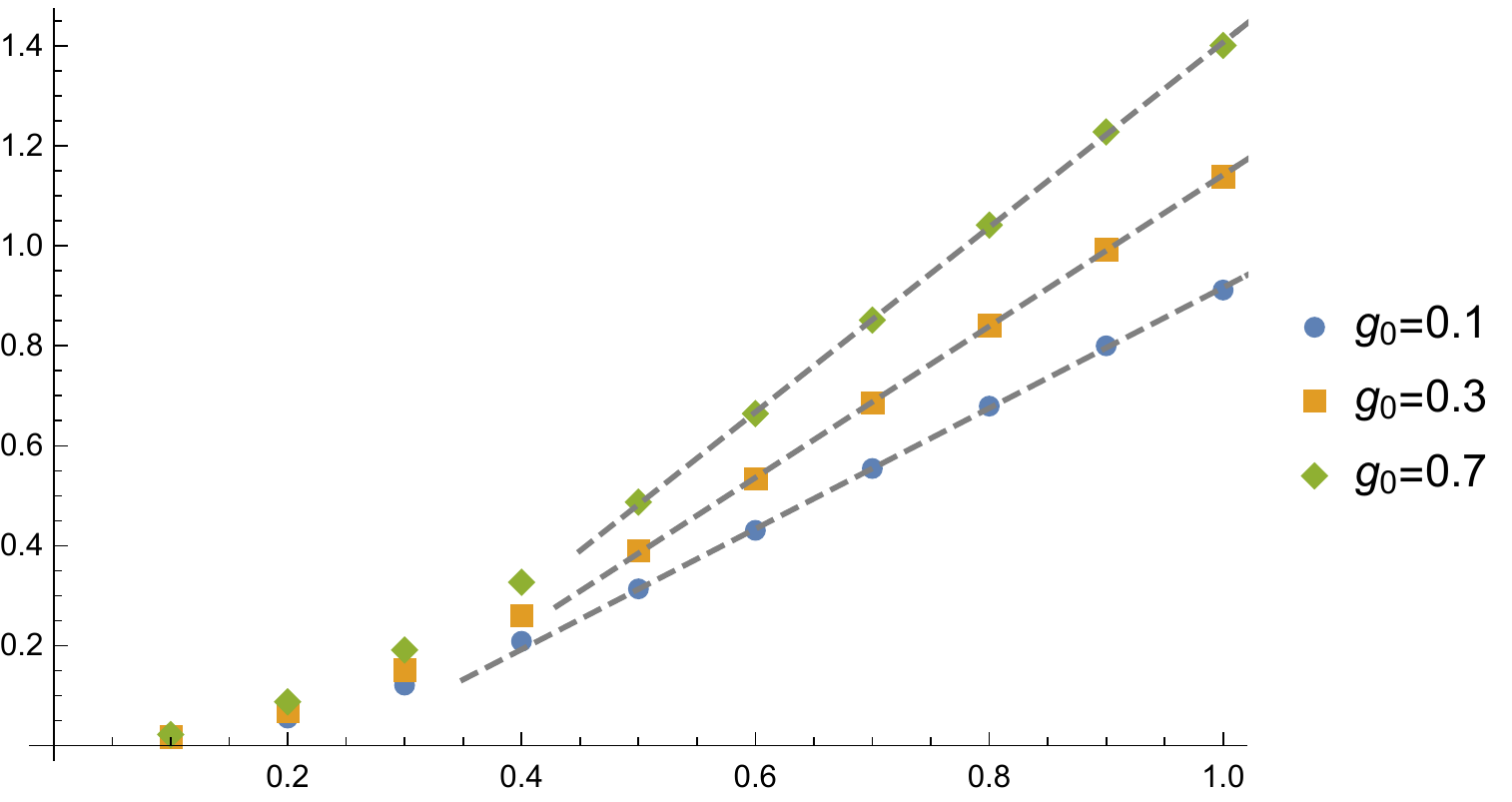}
}
\caption{K-complexity evolution in different time stages. When t<0.1, the K-complexity shows a quadratic growth. During $0.1<t<1$, K-complexity deviates from the quadratic growth and then grows linearly with time.}
\label{K-complexity growth}
\end{figure}

Thus, the K-complexity is expected to grow quadratically at the beginning. We show the early growth of K-complexity under different coupling constants in figure \ref{K-complexity growth}. As expected, in the early time stage, the K-complexity shows a quadratic growth behavior. The dash lines in the figure are fitting functions $C_K(t) = \gamma t^{2}$. The coefficients $\gamma$ are respectively $\gamma_{0.1} = 1.39$, $\gamma_{0.3} = 1.76$, $\gamma_{0.7} = 2.23$. Then in the period $0.1<t<1$, K-complexity starts to deviate from quadratic growth and turn into linear growth. The linear growth can be numerically fitted by $C_K(t) = \eta t + \xi$. The slop $\eta$ for different coupling constants are $\eta_{0.1} = 1.21$, $\eta_{0.3} = 1.51$, $\eta_{0.7} = 1.85$. From the analysis of Lanczos coefficients in section \ref{4.2}, this linear growth will last forever. 

As stated in section \ref{3.1}, the power-law growth of K-complexity corresponds to integrable systems, while the linear growth corresponds to chaotic systems. Therefore, it can be seen clearly that in the CHO model, there is a transformation from integrability to chaos. The transition time $0.1 < t_{0} < 1$. The exponential growth of K-complexity (\ref{250}) is also suppressed by this effect.

\section{Microcanonical K-complexity}\label{section5}
As we saw in the classical analysis, the energy of the system will affect its chaotic nature. In the thermal state, high-modes show more chaotic properties, while low-modes show more integrable properties. In this section, we will analyze the effect of different energies as well as different temperatures on the K-complexity and Lanczos coefficients.
\subsection{Contribution from different energy}
To investigate the contribution from different energy modes to the K-complexity. we change the Wightman inner product as follows
\begin{equation}\label{379}
    (A|B)_E \equiv \frac{1}{Z} \sum_{E_n = E} \langle n | e^{H \beta /2} A^\dagger e^{-H \beta /2} B | n \rangle
\end{equation}
where the summation means summing over all the states with degeneracy $E_n = E$. In this way, the original Wightman inner product can be expressed as
\begin{equation}
    (A|B) = \int dE \ \rho(E) e^{-\beta E} (A|B)_E
\end{equation}
where we defined $\rho(E) = \sum_{n}(E_n - E) $. Keeping the Krylov basis fixed, according to eq(\ref{amplitude}), our Krylov amplitude becomes
\begin{equation}
    \varphi_n(t,E) = i^{-n} (\mathcal{O}_n|\mathcal{O}(t))_E
\end{equation}
This is related to the original Krylov amplitude by
\begin{equation}
    \varphi_n(t) =\int dE \ \rho(E) e^{-\beta E} \varphi_n(t,E)
\end{equation}
With this definition, we can extract the contribution from different energy sectors to the Krylov amplitude $\varphi_n(t)$.

\begin{figure}[htbp] 
\centering 
\includegraphics[width=0.6\textwidth]{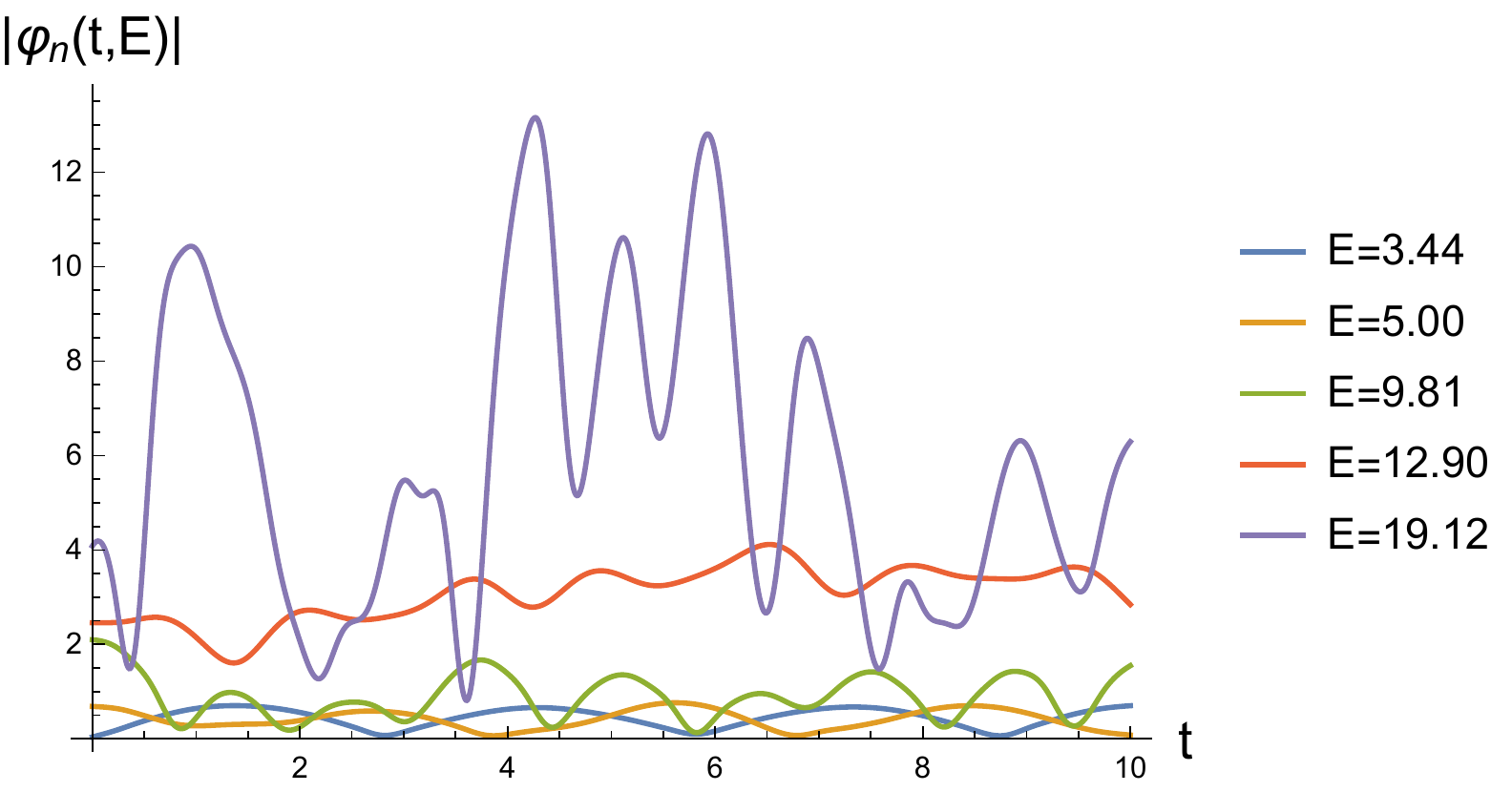} 
\caption{The norm of energy-dependent Krylov amplitude $\varphi_n(t,E)$ vs evolution time $t$. Here we choose $n=10$ and five different values of energy $E=3.44,E=5.00,E=9.81,E=12.90,E=19.12$. At low energy sectors, $|\varphi_n(t,E)|$ has a strong periodicity. While for high-energy, the periodicity of $|\varphi_n(t,E)|$ is broken and the oscillations become random and erratic. on the other hand, high-modes of $|\varphi_n(t,E)|$ have larger values than low-modes, which means it can contribute more to the K-complexity.} 
\label{fig421} 
\end{figure}

We show in figure \ref{fig421} the evolution of $|\varphi_n(t,E)|$ for different energies $E$ in the CHO model with $g_0=0.1$ and $n=10$. It can be seen that after lifting the suppression of the Boltzmann factor $e^{-\beta E}$, the high-energy modes have a significantly larger contribution to $\varphi_n(t)$. Notice that when E is small, $\varphi_n(t,E)$ has a distinct periodicity, which is a characteristic of the integrable system. In other words, the effect of the non-linear term in the CHO model is more on the higher energy modes. Therefore, when the energy of the system is elevated, it deviates from the integrable system and shows stronger chaotic features.

\subsection{Energy and temperature dependence of Lanczos coefficients}

In the classical limit, the Lanczos coefficients growth rate $a$ can be calculated by the singularity of the analytically continued auto-correlation function $C(t)$. Consider a classical orbit $E = H(x,y,\frac{\partial x}{\partial t},\frac{\partial y}{\partial t})$. Assume the equations of motion have the solution $x(t)=F_E(t)$, and inversely, $t(x)=F_E^{-1}(x)$. $x$ is analytic for all $t\in \mathbb{R}$. So the singularity in $x(t)$ has the imagination $\sigma_* = -i F_E^{-1}(\infty)$. So, $x(t)$ is analytic in the strip $\{t:\Im(t) \leq \sigma_* \}$. Thus, the analytic region of auto-correlation function $C(t) = \langle x(t) x(0) \rangle$ can be determined by the singularity in $x(t)$. If we replace the ensemble average by the time average $C(t) = \frac{1}{T}\int x(s+t/2)x(s-t/2)ds$, then $C(t)$ is analytic in the strip $\{t:\Im(t) \leq  2 \sigma_* \}$. According to \cite{Avdoshkin:2019trj}, $2\sigma_*$ is related to the Lanczos coefficients growth rate $a$ by $a = \frac{\pi}{4 \sigma_*}$. 

In the high energy limit, the harmonic potential $\frac{\omega^2}{4}(x^2+y^2)$ can be ignored. So, the remaining energy is $E = p_x^2+p_y^2+g_0 x^2 y^2$, which allows the scaling transformation, $x\to \Lambda x, y\to \Lambda y, t\to \Lambda^{-1} t, E\to \Lambda^4 E$. Plugging the scaling relation to $\sigma_*$, it turns out that $\sigma_*$ scales as $\sigma_* \propto E^{-1/4}$. Therefore, the growth rate $a$ scales as $a= \frac{\pi}{4 \sigma_*} \propto E^{1/4}$. In the low energy limit, the harmonic potential dominates, so $\sigma_*$ doesn't scale with the energy. 

\begin{figure}[htbp]
\centering 
\includegraphics[width=0.6\textwidth]{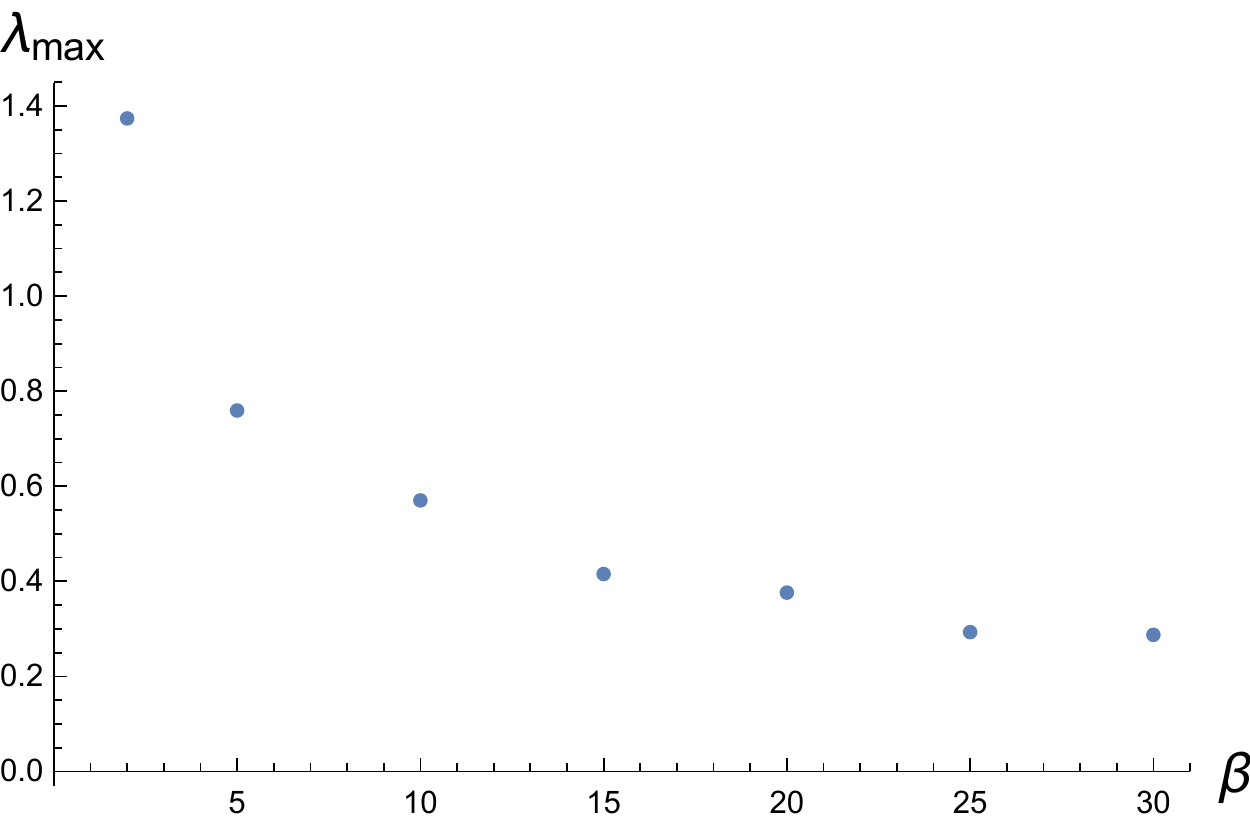} 
\caption{Numerical result for the maximum bound of quantum Lyapunov exponent $\lambda_{max} <2a$ in different temperature. The horizontal coordinate denotes the inverse temperature $\beta$.} 
\label{fig431} 
\end{figure}

In the thermal K-complexity, the contribution from different energy regions is suppressed by the Boltzmann factor $e^{-\beta E_n}$. This means that increasing temperature will enhance the contribution from higher energy modes. We show in figure \ref{fig431} the upper bound of quantum Lyapunov exponent $\lambda_{max} = 2a$ for different temperatures when the coupling constant $g_0=0.1$.

\section{Conclusion and discussion}\label{section6}

In this paper, we investigated the Krylov complexity and associated Lanczos coefficients growth in the SU(2) Yang-Mills theory, which can be reduced to a nonlinearly coupled harmonic oscillators model. We developed a method to numerically calculate Krylov complexity in this chaotic model. We computed the growth patterns of K-complexity and Lanczos coefficients for different coupling constants. We observed a transition from integrability to chaos in the CHO model. The K-complexity grows quadratically in the early stage, which is characteristic of integrable systems. After the transition time $t_0$, K-complexity grows linearly, which is characteristic of chaotic systems.

While for Lanczos coefficients, we found that it linearly increases with $n$ when $n$ is small and then enters the saturation plateau, which is consistent with the universal hypothesis in \cite{Parker_2019}. It was conjectured that the quantum Lyapunov exponent $\lambda$ is upper bounded by the slope of the linear growth of the Lanczos coefficients $a$, i.e.,  $\lambda < 2a$. Using this relation, we found the upper bound of the quantum Lyapunov exponent in the CHO model at different temperatures and coupling constants.

By analyzing the microcanonical K-complexity, we extracted the contributions from different energy sectors to the K-complexity and Lanczos coefficients. We found that higher energy modes have a greater contribution to the growth of K-complexity, which means that the higher the energy, the more the system shows the nature of quantum chaos. This is consistent with our classical analysis. Classically, by analyzing the singularity of the auto-correlation function $C(t)$, we obtained that the bound for Lyapunov exponent $\lambda_{max}$ is proportional to $E^{1/4}$ in the high-energy limit. In the thermal state, the energy of the system is closely related to the temperature. We calculated the effect of different temperatures on Lanczos coefficients. In addition, we also numerically computed the spectral form factor in the CHO model. We found a behavior similar to that of the random matrix theory, which indicates there might be a random matrix description of SU(2) Yang-Mills theory.

In conclusion, in SU(2) Yang-Mills theory, Krylov complexity is an interesting observable which encodes many features of quantum chaos. This may have implications for a good definition of quantum chaos. Also, the application of Krylov complexity to strongly interacting field theories may advance the search for its gravity dual. Although our study focuses on the two-dimensional CHO model, our approach can be generalized to higher dimensions. For future research, it will be interesting to see whether Krylov complexity can be computed in more general quantum chaotic systems and quantum field theories.

\bibliographystyle{JHEP}
\bibliography{reference}

\end{document}